\documentclass[preprint, nofootinbib]{revtex4-1}

\usepackage{amsmath}
\usepackage{amsfonts,amssymb}
\usepackage{hyperref}
\usepackage{graphicx}
\usepackage{natbib}
\usepackage[utf8]{inputenc}

\begin{document}

\title{Does the cosmological constant stay hidden?}
\author{Oliver F. Piattella}
\email{oliver.piattella@cosmo-ufes.org}

\affiliation{N\'ucleo Cosmo-UFES \& Department of Physics, Universidade Federal do Esp\'irito Santo, Avenida Fernando Ferrari 514, 29075-910 Vit\'oria, Esp\'irito Santo, Brazil}

\begin{abstract}
 We elaborate on the proposal of [Phys. Rev. Lett. 123 (2019) 13, 131302], about the possibility of hiding the cosmological constant in the complicated topology that one expects to exist at the Planck scale. We build a differential equation ruling the time evolution of $\langle K\rangle$, the spatial average of the expansion scalar. Supposing that the solution $\langle K\rangle = 0$ exists despite the presence of a large cosmological constant $\Lambda$, we show that such solution seems to be unstable.
\end{abstract}

\maketitle

\section{Introduction}

The cosmological constant is an ingredient of General Relativity (GR) which is both theoretically satisfactory, if one thinks about Lovelock's theorems \cite{Lovelock:1971yv, Lovelock:1972vz}, and observationally successful, if one looks at the $\Lambda$CDM model of cosmology and its effectiveness in describing our universe on the largest scales, especially from the point of view of its accelerated expansion discovered about two decades ago \cite{Schmidt:1998ys, Riess:1998cb, Perlmutter:1998np}. 

Despite these fulfillments, something with $\Lambda$ seems to be problematic. We expect, on the fundamental ground of the equivalence principle, that the minima of potentials, to which the quantum fields of our world have settled, provide contributions to the cosmological constant. While we could rescale these minima to zero, thereby avoiding any extra contribution to $\Lambda$, this cannot be done both before and after a phase transition. For the spontaneous symmetry breaking of the Higgs mechanism, for example, the minimum of the Higgs potential shifts of an amount of $10^8$ GeV$^4$, whereas the value of the observed energy density associated to $\Lambda$ is $10^{-47}$ GeV$^4$. So, why should the minimum settle within $10^{-47}$ GeV$^4$ upon a ``jump'' of 55 orders of magnitude larger, is a mystery. Attempts to solve the latter include, for example, the possibility of a running of the cosmological constant \cite{Shapiro:1999zt, Shapiro:2009dh}. 

From a quantum field theory perspective, also the zero-point energy of quantum fields should gravitate,\footnote{Historically, Zel'dovich \cite{Zeldovich:1967gd} was the first to employ vacuum energy as an ingredient of a cosmological model which could explain some astronomical observations.} thereby behaving as a cosmological constant. It is frequently speculated that this contribution is huge, so that no universe in which structures form would be possible \cite{barrow1982isotropy, barrow1986anthropic, Weinberg:1988cp, Martin:2012bt}.\footnote{It seems that this had been realised already by Pauli in the 1920s, see e.g. \cite{Straumann:2002he}, when he estimated the influence of the zero-point energy of the radiation field (with a cutoff at the classical electron radius) on the radius of the universe, and came to the conclusion that it ``could not even reach to the moon''.} On the other hand, no one really knows how to compute the zero-point energy in the realistic case of interacting quantum fields and in presence of spacetime curvature, so we cannot really say much in this respect. Perhaps, as speculated in Ref.~\cite{Holland:2013xya}, the result is exponentially suppressed, so no problem arises from this side. 

 The problem of explaining why vacuum energy should not gravitate is nowadays called \textit{old cosmological constant problem}. It is typically framed in the semiclassical approach to GR, and probably its solution demands a clearer comprehension of whether and how vacuum energy gravitates and a clearer understanding of how to implement quantum effects in a theory of gravitation. Moreover, observation presents us with the fact that $\Lambda$ has an energy density comparable to that of matter, so a \textit{new cosmological constant problem} has arisen, related with the famous \textit{coincidence problem} of cosmology, which requires the explanation of how vacuum energy can be tuned in that way \cite{Weinberg:2000yb}.  

In this paper we are interested in one recent proposal, which addresses the old formulation of the cosmological constant problem, that of Ref. \cite{Carlip:2018zsk} (see also Ref.~\cite{Carlip:2019mba}). Here it is put forward the possibility that even if a huge $\Lambda$ does exist at the Planck scale, it can nevertheless be averaged to vanishingly small values on macroscopic scales. This averaging to zero depends on the foliation chosen in a 3 + 1 decomposition of spacetime, in particular on the lapse function $N$, but there are infinite possible choices of $N$ that allow to ``hide'' a huge $\Lambda$ within the averaging process which eventually result in a vanishingly small averaged expansion scalar $\langle K\rangle$. Put in other words, a huge $\Lambda$ is hidden in the foamy nature of spacetime, i.e. Wheeler's famous ``spacetime foam'' \cite{Wheeler:1955zz}. The proposal of Ref. \cite{Carlip:2018zsk} has received some attention and debate, see e.g. Refs.~\cite{Wang:2019wwg, Carlip:2019aaz, Wang:2020cvm, Carlip:2020yfs}.

Here, we are mostly interested in the time evolution of $\langle K\rangle$, i.e. whether the ``hiding'' is preserved with time. In Ref.~\cite{Carlip:2018zsk} the focus is on the Cauchy surface of the initial values on which it is argued that the averaged expansion scalar and its time-derivatives can be made vanishing for infinite choices of $N$ and its time-derivatives. Here we follow a different path: we derive an evolution equation for $\langle K\rangle$ and analyse the stability of its $\langle K\rangle = 0$ solution. We find that, despite the ingenuity of the idea of the hiding, the solution $\langle K\rangle = 0$ seems to be unstable. It must be stressed that already in Ref.~\cite{Carlip:2018zsk} a caveat is given about the classical evolution of $\langle K\rangle$, which is valid only for short times due to the inevitable appearance of singularities.

\section{Initial values formulation of General Relativity and the averaging of scalar quantities}

In this section we gather the relevant equations of our framework, i.e. the initial value formulation of General Relativity (GR) \cite{Arnowitt:1962hi, Ellis:1971pg, Buchert:1999er, Buchert:2001sa, Carlip:2018zsk, Buchert:2019mvq}. Consider a 3+1 foliation of spacetime into a family of spacelike hypersurfaces $\Sigma_t$ orthogonal to a time-like vector $n_\mu$, normalised as $n_\mu n^\mu = -1$. From a general metric $g_{\mu\nu}$ one can define the projector $h_{\mu\nu}$ as:
\begin{equation}
	h_{\mu\nu} = g_{\mu\nu} + n_\mu n_\nu\;,
\end{equation} 
where the plus sign comes from having chosen the positive signature for $g_{\mu\nu}$. The metric $g_{\mu\nu}$ can be then decomposed as follows:
\begin{equation}\label{decomposedmetric}
	ds^2 = -(N^2 - N_iN^j)dt^2 + 2N_idx^idt + g_{ij}dx^idx^j = -N^2dt^2 + g_{ij}(dx^i + N^idt)(dx^j + N^jdt)\;,
\end{equation}
where $N(t,\mathbf x)$ is the lapse function, $N^i(t,\mathbf x)$ is the shift function and $g_{ij} = h_{ij}$. The extrinsic curvature of $\Sigma_t$ is the symmetric part of the projection on $\Sigma_t$ of the covariant derivative of $n_\mu$:
\begin{equation}
	K_{\rho\sigma} := n_{\mu;\nu}h^\mu{}_{(\rho} h^\nu{}_{\sigma)}\;.
\end{equation}
We define it with the plus sign here, following Ref.~\cite{Carlip:2018zsk}, instead of the minus one, used e.g. in Ref.~\cite{Buchert:2001sa}. The antisymmetric part of the projected covariant derivative of $n_\mu$ is the so-called twist and by construction it is absent here because the very possibility of performing a $3 + 1$ foliation depends precisely on the vanishing of the twist. 

The Einstein equations can be rewritten on the foliation as two constraints and two evolution equations for $g_{ij}$ and $K^i{}_j$. The Hamiltonian constraint is:
\begin{equation}
	R - K^i{}_jK^j{}_i + K^2 - 2\Lambda = 16\pi GT_{\mu\nu}n^\mu n^\nu\;, 
\end{equation}
where $T_{\mu\nu}$ is the matter energy-momentum tensor and we have also included the cosmological constant $\Lambda$. The momentum constraint is:
\begin{equation}
	D_i(K^i{}_{j} - \delta^i_jK) = 8\pi G T_{\mu\nu}n^\mu h^\nu{}_i\;,
\end{equation}
where $D_i$ denotes the covariant derivative with respect to the spatial metric $h_{ij} = g_{ij}$.

The evolution equation for the first fundamental form (i.e. the spatial metric) is:
\begin{equation}
	\frac{1}{N}\dot{g}_{ij} = 2K_{ij} + \frac{1}{N}(D_jN_{i} + D_iN_{j})\;,
\end{equation}
where the dot denotes derivation with respect to $t$. The evolution equation for the second fundamental form (i.e. the extrinsic curvature tensor) is:
\begin{eqnarray}
	\frac{1}{N}\dot{K}^i{}_j = -R^i{}_j - KK^i{}_j + \delta^i_j\Lambda + \frac{D^iD_jN}{N} + \frac{1}{N}(K^i{}_kD_jN^k - K^k{}_jD_kN^i + N^kD_kK^i{}_{j}) \nonumber\\
	+ 8\pi G\left[\mathcal S^i{}_j + \frac{1}{2}\delta^i_j(\epsilon - \mathcal S^k{}_k)\right]\;, \qquad \mathcal S_{ij} := T_{\mu\nu}h^\mu{}_i h^\nu{}_j\;,
\end{eqnarray}
where $R^i{}_j$ is the Ricci scalar of the hypersurface $\Sigma_t$.

From the two evolution equations we can infer how the determinant $g \equiv \det(g_{ij}) $ evolves with time:
\begin{equation}\label{gdoteq}
	\frac{1}{N}\dot{g} = 2g\left(K + \frac{1}{N}D_kN^k\right)\;,
\end{equation}
and the Raychaudhuri equation for the expansion scalar $K$, which is the trace of the extrinsic curvature:
\begin{equation}\label{Kdoteq}
	\frac{1}{N}\dot{K} = -R - K^2 + 3\Lambda + \frac{D^kD_kN}{N} + \frac{1}{N}N^kD_kK + 4\pi G(3T_{\mu\nu}n^\mu n^\nu - \mathcal S^k{}_k)\;.
\end{equation}

\subsection{Averaging} 

The main result of Ref.~\cite{Carlip:2018zsk} is to show that $\langle K \rangle$ is vanishingly small despite the presence of a very large $\Lambda$. In order to do that, the argument is to prove that if $\langle K \rangle = 0$ as initial condition on a certain $\Sigma_t$, then also all the time-derivatives of $\langle K \rangle$ are zero on the same $\Sigma_t$ and thus $\langle K \rangle$ is zero at all times. This can be achieved by suitably choosing $N$ and its time-derivatives on $\Sigma_t$. This serves to prove that even if a very large cosmological constant $\Lambda$ exists, it is absent (hidden) in the macroscopic average of the expansion $\langle K \rangle$ for many observers, i.e. for many choices of $N$. Note that ``macroscopic'' is intended here as ``much larger than the Planck size, but not as large as the size of the universe''.

Averaging in GR is an open problem \cite{Clarkson:2011zq} but if we restrict ourselves to scalar quantities, such as $K$, then a natural definition is the following \cite{Buchert:1999er, Buchert:2001sa}:
\begin{equation}\label{average}
	\langle X\rangle_{\mathcal D} = \frac{1}{V_{\mathcal D}}\int_{\mathcal D}X\sqrt{g}d^3x\;, \qquad V_{\mathcal D}= \int_{\mathcal D}\sqrt{g}d^3x\;.
\end{equation}
The average depends of course on the portion of space $\mathcal D$ over which it is performed. We avoid the subscript $\mathcal D$ from now on.

\section{The evolution equation for $\langle K\rangle$}

Let us neglect matter with respect to the cosmological constant $\Lambda$ (which might be thought of as the effective one, i.e. also incorporating the zero-point energy of the quantum fields, and thus it might possibly be very large). The average $\langle K \rangle$ is, using the definition \eqref{average}:
\begin{equation}\label{Kaverage}
	\langle K\rangle = \frac{1}{V}\int K\sqrt{g}d^3x\;,
\end{equation}
so, by taking the time derivative of this and employing Eq.~\eqref{Kdoteq}, one gets:
\begin{eqnarray}\label{avKdoteq}
	\langle K \rangle\dot{} = -\frac{\dot V}{V}\langle K \rangle + \frac{1}{V}\int[\dot K + K\dot g/(2g)]\sqrt{g}d^3x = \nonumber\\
	-\frac{\dot V}{V}\langle K \rangle + \frac{1}{V}\int[N(-R + 3\Lambda) + D^kD_kN + N^kD_kK + KD_kN^k]\sqrt{g}d^3x\;.
\end{eqnarray}
The derivative of the volume is obtained by making use of Eq.~\eqref{gdoteq}:
\begin{equation}\label{Vdoteq}
	\frac{\dot V}{V} = \frac{1}{V}\int[\dot g/(2g)]\sqrt{g}d^3x = \frac{1}{V}\int (NK + D_kN^k)\sqrt{g}d^3x\;.
\end{equation}
Note that the computation of the time derivative of the volume is not necessary in Ref.~\cite{Carlip:2018zsk} because of the initial condition chosen such that $\langle K \rangle = 0$. Combining Eqs.~\eqref{avKdoteq} and \eqref{Vdoteq} we then obtain the following differential equation for $\langle K \rangle$:
\begin{eqnarray}\label{Kdoteq2}
	\langle K \rangle\dot{} = -(\langle NK \rangle + \mathcal B_1)\langle K \rangle - \langle NR \rangle + 3\langle N \rangle\Lambda + \mathcal B_2 + \mathcal K\;,
\end{eqnarray}
where $\mathcal B_{1,2}$ are boundary terms depending only on the shift and lapse functions, respectively:
\begin{equation}
	\mathcal B_1 := \frac{1}{V}\int D_kN^k\sqrt{g}d^3x\;, \quad \mathcal B_2 := \frac{1}{V}\int D^kD_kN\sqrt{g}d^3x\;,
\end{equation}
and $\mathcal K$ is a boundary term involving $K$ itself:
\begin{equation}
	\mathcal K := \frac{1}{V}\int D_k(KN^k)\sqrt{g}d^3x\;.
\end{equation}
If we denote as $v_k$ the normal vector to the boundary $\partial\mathcal D$, we have that the boundary terms are surface integrals of the quantities $v_kN^k$, $v^kD_kN$ and $Kv_kN^k$. Even if we fix $N$ and $N^k$, still we have the freedom of choosing an arbitrary $\mathcal D$, with an arbitrary boundary $\partial\mathcal D$, which amounts to the freedom of choosing an arbitrary vector field $v_k$. Therefore, at least one of the boundary terms can be looked as an arbitrary function, even if we have already fixed $N$ and $N^k$.

The term $\langle NK \rangle$ prevents us from having a closed equation for $\langle K \rangle$. However, we can overcome this hurdle by using the following lemma, proved in Refs.~\cite{Buchert:1999er, Buchert:2001sa}. For a generic quantity $\Psi$ and the average defined in Eq.~\eqref{average} one has:
\begin{equation}
	\langle \Psi \rangle\dot{} = \langle \dot\Psi \rangle + \langle NK\Psi\rangle - \langle NK \rangle\langle \Psi \rangle\;,
\end{equation}
Choosing then $\Psi = 1/N$, one gets:
\begin{equation}
	\langle 1/N \rangle\dot{} = -\langle \dot N/N^2 \rangle + \langle K\rangle - \langle NK \rangle\langle 1/N \rangle\;,
\end{equation}
from which we obtain:
\begin{equation}
	\langle NK \rangle = \frac{\langle K \rangle}{\langle 1/N \rangle} - \frac{\langle 1/N \rangle\dot{} + \langle \dot N/N^2 \rangle}{\langle 1/N \rangle}\;.
\end{equation}
Substituting this result into Eq.~\eqref{Kdoteq2} we obtain a closed differential equation for $\langle K \rangle$:
\begin{equation}\label{Kdoteq3}
	\langle K \rangle\dot{} = -f_2\langle K \rangle^2 + f_1\langle K \rangle + f_0\;,
\end{equation}
i.e. a Riccati-type equation, with:
\begin{equation}\label{defsfgh}
	f_2 := \frac{1}{\langle 1/N \rangle}\;, \quad f_1 := \frac{\langle 1/N \rangle\dot{} + \langle \dot N/N^2 \rangle}{\langle 1/N \rangle} - \mathcal B_1\;, \quad f_0 = -\langle NR\rangle + 3\langle N \rangle\Lambda + \mathcal B_2 + \mathcal K\;.
\end{equation}
Since:
\begin{equation}
	f_1 = -\frac{\dot f_2}{f_2} + f_2\langle \dot N/N^2 \rangle - \mathcal B_1\;,
\end{equation}
we can cast Eq.~\eqref{Kdoteq3} as follows:
\begin{equation}\label{Kdoteq4}
	(f_2\langle K \rangle)\dot{} = -(f_2\langle K \rangle)^2 + (f_2\langle \dot N/N^2 \rangle - \mathcal B_1)(f_2\langle K \rangle) + f_2f_0\;,
\end{equation}
using $f_2\langle K \rangle$ as the unknown function.

Since $N > 0$, i.e. the lapse function is strictly positive because an arrow of time is established, then $f_2 > 0$, provided $\sqrt{g} > 0$. If singularities develop, for example due to the gluing technique which allows us to choose a vanishing initial $\langle K \rangle$, see e.g. \cite{Burkhart:2019zbx}, $\sqrt{g}$ might diverge somewhere in the averaging region badly enough to make, despite the integration, $\langle 1/N \rangle$ diverging and thus $f_2$ to vanish. We do not consider this possibility here and simply assume $f_2 > 0$ from now on.

\subsection{The stability condition in the case of negligbile boundary terms}

Let us now focus on the simplest case, in which we neglect the  boundary terms $\mathcal B_{1,2}$ and $\mathcal K$. The latter makes Eq.~\eqref{Kdoteq4} especially tricky since it contains $K$ itself, so we assume the boundary terms, being surface terms, to be negligible with respect to the ``bulk'' terms. Incidentally, this assumption is equivalent to the one in which we neglect the shift, explicitly used in Ref.~\cite{Carlip:2018zsk}, and $\mathcal B_2$, also assumed in Ref.~\cite{Carlip:2018zsk}, but less explicitly.

Thanks to our assumptions we can cast Eq.~\eqref{Kdoteq4} as follows:
\begin{equation}\label{Kdoteq5}
	(f_2\langle K \rangle)\dot{} = -(f_2\langle K \rangle)^2 + f_2\langle \dot N/N^2 \rangle (f_2\langle K \rangle) + f_2(-\langle NR\rangle + 3\langle N \rangle\Lambda)\;.
\end{equation}
The solution $\langle K \rangle = 0$ exists if:
\begin{equation}\label{cond1}
	\langle NR\rangle = 3\langle N \rangle\Lambda\;.
\end{equation} 
The same condition is required in Ref.~\cite{Carlip:2018zsk}, after their Eq. (7), but only on the initial values hypersurface. Here we need it to hold true throughout the whole time evolution. This seems already somehow problematic because the time dependence of the right hand side of Eq.~\eqref{cond1} comes only from $N$ and $g$, whereas on the left hand side we have also the time dependence of $R$. Let us assume anyway, that such a choice is possible and $\langle K \rangle = 0$ is a solution. The question is now, is this solution stable?

It is not difficult to see from Eq.~\eqref{Kdoteq5}, with $\langle NR\rangle = 3\langle N \rangle\Lambda$, that $\langle K \rangle = 0$ is a stable solution if:
\begin{equation}\label{cond2}
	f_2\langle \dot N/N^2 \rangle < 0\;.
\end{equation} 
This can be achieved only if $\dot N < 0$, provided again that no singularities for which $g = 0$ develop. But since $N > 0$, $N$ cannot arbitrarily decrease. At a certain time $\bar t$ we expect $\dot N$ to vanish and possibly to change sign. When this happens, $\langle K \rangle$ would start to grow away from $\langle K \rangle = 0$. However, the time $\bar t$ can be infinite if $N$ tends asymptotically to zero from above, i.e. it has a time behaviour such as $e^{-\lambda t}$, for example, with $\lambda >0$. Note that, if we choose the following form for $N$:
\begin{equation}\label{varsep}
	N(t,\mathbf x) = T(t)X(\mathbf x)\;,
\end{equation}
i.e. if we assume variable separation, then in condition \eqref{cond1} the time dependence of $N$, i.e. $T$, cancels out. Thus, we have on both sides the same time dependence of the averages induced by $\sqrt{g}$ and, moreover, on the left hand side the extra time dependence due to $R$. This might turn problematic the fulfilment of such condition. Of course, $R$ constant would solve this particular problem, but then we would have $R = \Lambda$ in order to have a vanishing solution for $\langle K\rangle$ and such a huge spatial curvature is already ruled out by observation. Note also that, even if $N$ had a behaviour such as $e^{-\lambda t}X(\mathbf x)$, with $\lambda >0$, the proper time corresponding to $t = \infty$ would be:
\begin{equation}
	\tau = X(\mathbf x)\int_0^\infty e^{-\lambda t}dt = X(\mathbf x)\frac{1}{\lambda}\;,
\end{equation}
i.e. finite, with values depending on the spatial point considered. So, $N$ goes to zero in a finite proper time.

\subsection{The explicit solution}

We can also write down an explicit, though rather formal, solution of Eq.~\eqref{Kdoteq5} by exploiting a very useful property of Riccati equations, which allows us, if we know a particular solution say $\mathcal F$, to write down a general solution as follows:
\begin{equation}\label{gensol}
	f_2\langle K \rangle = \mathcal F + \frac{\Phi(t)}{C + \int^t dt'\Phi(t')}\;, \qquad \Phi = \exp\left\{\int^tdt'\left[-2\mathcal F(t') + f_2\langle \dot N/N^2 \rangle\right]\right\}\;,
\end{equation}
where $C$ is some integration constant. For the particular solution $\langle K \rangle = 0$, whose existence we have assumed, the general solution becomes then:
\begin{equation}\label{solhnull}
	\langle K \rangle = \frac{1}{f_2}\frac{\Phi(t)}{C + \int^t dt'\Phi(t')}\;, \qquad \Phi = \exp\left(\int^tdt'f_2\langle \dot N/N^2 \rangle\right)\;.
\end{equation}
A reflection of the above mentioned instability is seen in the denominator of Eq.~\eqref{solhnull}. Indeed, $1/C$ is the initial value of $f_2\langle K \rangle$. So, if $f_2\langle K \rangle$ is vanishingly small but not with fixed sign, then $C$ is large and positive or negative. When $C < 0$, the denominator:
\begin{equation}
	C + \int^t dt'\Phi(t')\;,
\end{equation}
might diverge, because $\int^t dt'\Phi(t')$ is always positive. In particular, we expect this to happen if $f_2\langle \dot N/N^2 \rangle > 0$, because in this case $\Phi$ is a growing function. If $C > 0$ there is no such divergence, but still $\langle K\rangle$ should in principle grow away from $\langle K\rangle = 0$ if $f_2\langle \dot N/N^2 \rangle > 0$, at least until the $-(f_2\langle K \rangle)^2$ term in the Riccati equation \eqref{Kdoteq5} dominates on the $f_2\langle \dot N/N^2 \rangle (f_2\langle K \rangle)$ one, in which case $\langle K\rangle$ starts again to decrease.

\subsection{Examples}

\begin{figure}[h]
	\includegraphics{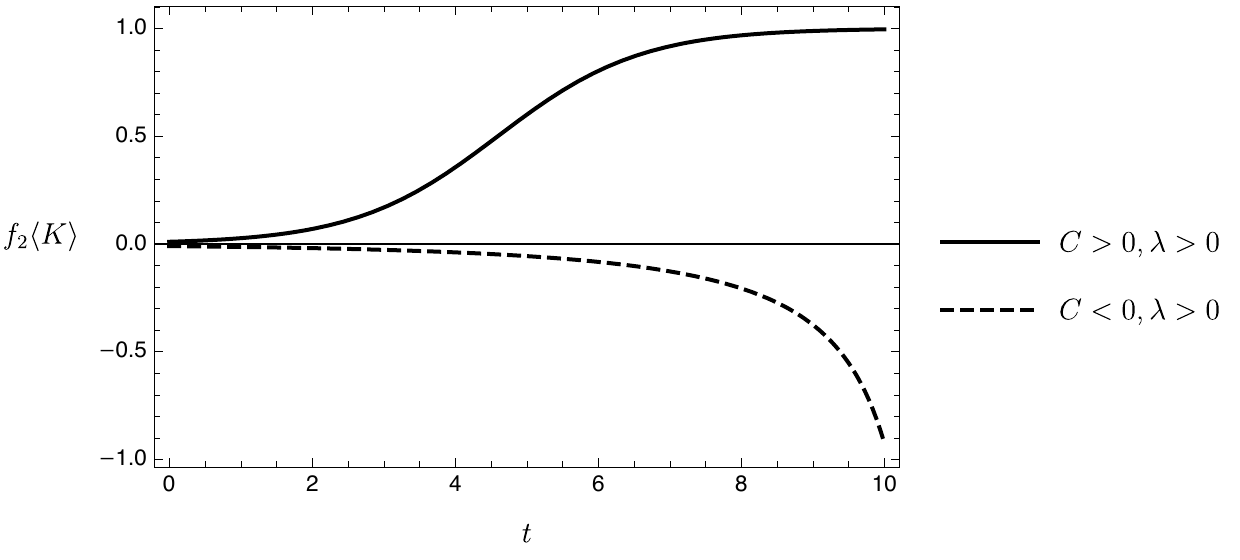}
	\caption{Evolution of $f_2$$\langle K \rangle$ for the cases 1 and 2 discussed in the text.}
	\label{Fig:plot}
\end{figure}

In this subsection we present a couple of concrete examples of the time evolution of $\langle K\rangle$. Let us assume first an exponential function for $\Phi$, i.e. $\Phi = \exp(\lambda t)$, with $\lambda$ positive or negative. Solution \eqref{solhnull} then becomes:
\begin{equation}
	\langle K \rangle = \frac{1}{f_2}\frac{e^{\lambda t}}{C + (e^{\lambda t} - 1)/\lambda}\;, 
\end{equation}
where we assume the initial time to be $t = 0$ and the initial value of $\langle K \rangle$ to be then $1/(f_{2}C)$, with $f_2$ evaluated at $t = 0$. We now must distinguish among 4 cases:

\begin{enumerate}
	\item If $C > 0$ and $\lambda > 0$, for sufficiently large times we have:
\begin{equation}
	\langle K \rangle \sim \frac{\lambda}{f_2}\;. 
\end{equation}
We do not know how $f_2$ evolves with time, but it is reasonable to assume that it is a limited function and thus the above solution tells us that $\langle K \rangle$ varies asymptotically in a certain range of values, none of which is vanishing.

\item If $C < 0$ and $\lambda > 0$, our solution can be written as:
\begin{equation}
	\langle K \rangle = \frac{1}{f_2}\frac{e^{\lambda t}}{(e^{\lambda t} - 1)/\lambda - |C|}\;. 
\end{equation}
As anticipated:
\begin{equation}
	\langle K \rangle \to -\infty\;, \quad \mbox{for}\quad t \to \frac{1}{\lambda}\ln(1 + \lambda|C|)\;,
\end{equation}
i.e. $\langle K \rangle$ diverges in a finite time interval.

\item If $C > 0$ and $\lambda < 0$, we have:
\begin{equation}
	\langle K \rangle = \frac{1}{f_2}\frac{e^{-|\lambda| t}}{C + (1 - e^{-|\lambda|t})/|\lambda|}\;. 
\end{equation}
This case seems to work fine because $\langle K \rangle$ goes to zero for $t \to \infty$. However, we have seen that $\lambda < 0$ demands $\dot N < 0$ and thus during the time evolution we must expect $N \to 0$. When this happens in a finite time, then $f_2 \to 0$ and thus $\langle K \rangle$ blows up. If $N \to 0$ instead for $t \to \infty$, we have then an indeterminate $0/0$ expression for $\langle K \rangle$ that we could resolve only knowing the details of the foliation. It might be that this case works, but still there is the problem of $N \to 0$. 

\item If $C < 0$ and $\lambda < 0$, we have:
\begin{equation}
	\langle K \rangle = \frac{1}{f_2}\frac{e^{-|\lambda| t}}{(1 - e^{-|\lambda|t})/|\lambda| - |C|}\;. 
\end{equation}
In this case, if $|C| < 1/\lambda$, the same analysis of case 2 applies, since the denominator of the expression above goes to zero in a finite time. If, on the other hand, $|C| > 1/\lambda$, then the same analysis of point 3 applies.
\end{enumerate}
In Fig. \ref{Fig:plot} we display the qualitative behaviour of the evolution of $f_2$$\langle K \rangle$ for the cases 1 and 2 discussed above.

Another interesting application is to assume variable separation, cf. Eq.~\eqref{varsep}, despite the fact that this choice might be inadequate for the fulfilment of condition \eqref{cond1}. In this case one finds:
\begin{equation}
	\langle K \rangle = \frac{\langle 1/X\rangle}{C + \int^t dt'T(t')}\;.
\end{equation}
Introducing the proper time we obtain:
\begin{equation}
	\langle K \rangle = \frac{\langle 1/X\rangle}{C + \tau/X}\;.
\end{equation}
Note that we have here also the time-dependence of $\langle 1/X\rangle$, which is induced by that of $\sqrt{g}$, which is unknown to us. This time dependence might not be well-behaved if singularities develop, but if $\langle 1/X\rangle$ does not diverge, then for $\tau \to \infty$ we do have $\langle K \rangle \to 0$ if $C > 0$ (for $C < 0$ we incur again into a divergence). 

\subsection{A different variable, closely related to the Hubble parameter}

Note that, according to Ref.~\cite{Buchert:1999er}, the ratio:
\begin{equation}\label{avNK}
	\frac{\dot V}{V} = \langle NK\rangle\;,
\end{equation}
where we have neglected the shift, cf. Eq.~\eqref{Vdoteq}, can be interpreted as an averaged Hubble factor and therefore one might argue that $\langle NK\rangle$ is the quantity we should focus on, rather than $\langle K\rangle$. It is not difficult to find, neglecting boundary terms:
\begin{eqnarray}\label{NKevoeq}
	\langle NK\rangle\dot{} = -\langle NK\rangle^2 + \langle \dot N/N\rangle \langle NK\rangle\nonumber\\ 
	+ \left(\langle \dot N/N^2\rangle\dot{} + \langle(\dot N/N^2)\dot{}\rangle + 3\langle N^2\rangle\Lambda - \langle N^2R\rangle + \langle ND^kD_kN\rangle\right)\;, 
\end{eqnarray}
as the evolution equation for $\langle NK\rangle$, whose structure is similar to that of the evolution equation for $\langle K\rangle$ and to which a similar analysis applies. An advantage of Eq.~\eqref{NKevoeq} over Eq.~\eqref{Kdoteq5} is that we do not have to worry about $f_2$ being strictly positive, but ``just'' on having:
\begin{equation}
	\langle \dot N/N^2\rangle\dot{} + \langle(\dot N/N^2)\dot{}\rangle + 3\langle N^2\rangle\Lambda - \langle N^2R\rangle + \langle ND^kD_kN\rangle = 0\;,
\end{equation}
in order to guarantee the $\langle NK\rangle = 0$ solution. Then, the same analysis presented above applies to this case.

\section{Discussion and conclusions}

In this paper we have investigated further the proposal of Ref.~\cite{Carlip:2018zsk}, about hiding a possibly huge cosmological constant into the foamy nature of spacetime at the Planck scale. Through a simple definition of average, cf. Eq.~\eqref{average}, we have built an evolution equation describing the time evolution of $\langle K\rangle$ and have analyzed the stability of its $\langle K\rangle = 0$ solution, provided that this exists. Unfortunately, it seems that such solution is unstable, because a necessary condition for its stability is $\dot N < 0$, which is not admissible throughout the whole evolution since $N > 0$. The fact that $N$ is required to decrease in order to have a stable solution might be an indication that singularities develop during the evolution, as already discussed in Ref.~\cite{Carlip:2018zsk}. There it also made the caveat that the $\langle K\rangle = 0$ solution found is short-lived. Through our investigation in the present paper, we confirm this conclusion. Therefore, at least for the very simple case considered, the hiding of $\Lambda$ is not preserved in time.   

\subsection*{Acknowledgements}

The author thanks S. Carlip and the anonymous referees for important suggestions and comments. This study was financed in part by the \emph{Coordena\c{c}\~ao de Aperfei\c{c}oamento de Pessoal de N\'ivel Superior} - Brazil (CAPES) - Finance Code 001. The author thanks the Alexander von Humboldt foundation for funding and the Institute for Theoretical Physics of the Heidelberg University for kind hospitality during part of the development of this project. The author also wishes to thank CNPq (Brazil) and FAPES (Brazil) for partial financial support.

\bibliographystyle{unsrturl}
\bibliography{CosmoC}

\end{document}